# Pion Gas Viscosity

Felipe J. Llanes-Estrada and Antonio Dobado.

The final stage of a Relativistic Heavy Ion Collision is an expanding meson gas. At low temperatures such gas needs to be composed mostly of pions, the lightest possible hadrons. The expansion is governed by a many particle flow, usually described as a hydrodynamic flow at zero baryon number. Viscous corrections might already be visible in the resulting HBT radii[1], and is in any case interesting to calculate the viscosity of a pion gas, because of the pions increasing interaction with energy. In a non relativistic, hard sphere gas, the viscosity is proportional to the mean particle velocity and mean free path, $\eta=(1/3) nmv\lambda$ and therefore to the root of the temperature after energy equipartition.

This is seen in recent calculations[2] based in the Uehling-Uhlenbeck equation (a quantum version of the Boltzmann equation for the distribution function) whose collision term was evaluated with a hard sphere interaction. At higher temperatures, comparable to the pion mass, the non-relativistic approximation breaks down and not only relativistic kinematics, but also a momentum dependent cross section must be used. We plot in figure 1 the result of an extension to relativistic kinematics of this approach, but maintaining the hard sphere approximation based on Weinberg´s theorem. This grossly overestimates the viscosity because the cross section is too small, and therefore the mean free path too large. Next we lift this approximation and use two simple parametrizations of the pion scattering phase shifts. The first of this[3] gives a much more reasonable viscosity of order 0.01 GeV3 which is plotted in figure 2. Notice this calculation disagrees by a factor of 2 with a previously published result[4] (probably due to their double-counting of identical particles in the final state by integrating over all solid angles ).

Finally we employ the phase shifts for pion scattering obtained with the Inverse Amplitude Method[5] and plot the viscosity in figure 3: it is of the same order of magnitude, but somewhat larger. To obtain the viscosity in all cases we equated the expressions for the stress tensor obtained from the hydrodynamic and the kinetical theory (the volume viscosity is known to be two orders of magnitude smaller than the shear viscosity):

The resulting expression is a one dimensional integral performed numerically. The most difficult part of the calculation (soon be reported at length) is to solve the kinetic equation for the distribution function. We profit from the 1917 Chapman-Enskog expansion around the equilibrium (Bose-Einstein) distribution function and approximately solve for the deviations out of equilibrium $f=f_0+\delta f$. The Uehling-Uhlenbeck equation

$Df/Dt = C[f]$ where $C[f]$ is the collisional term can then be linearized on the perturbation. The only difficulty is to evaluate the collisional term, which involves a multidimensional integral. To calculate it we employ the algorithm Vegas[6]. The results are checked against the non relativistic limit and comparing to a similar approach in ref. [4].ç

*Acknowledgements: The authors thank S. N. Santalla and especially F. J. Fernandez for extensive checks. Work supported by grants FPA2000-0956 and BFM 2002-01003.*

*Figure Captions:*

*Figure 1.- Viscosity as a function of temperature and fugacity calculated for a hard sphere gas (with the Weinberg scattering lengths). Units are GeV.*

*Figure 2.- Viscosity using the phase shifts from ref. 3. Units are GeV.*

*Figure 3.- Viscosity employing the IAM. Units are GeV.*

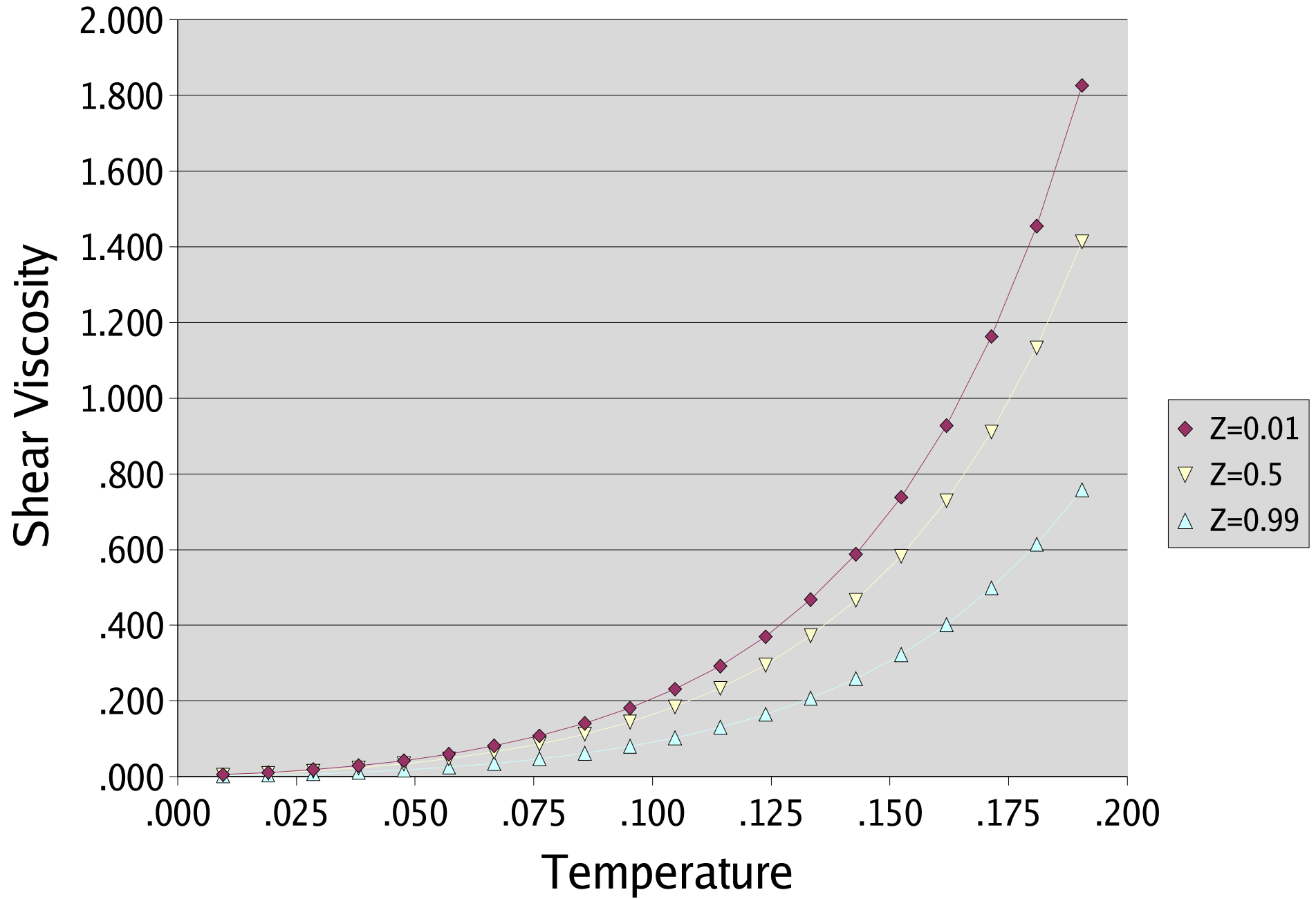

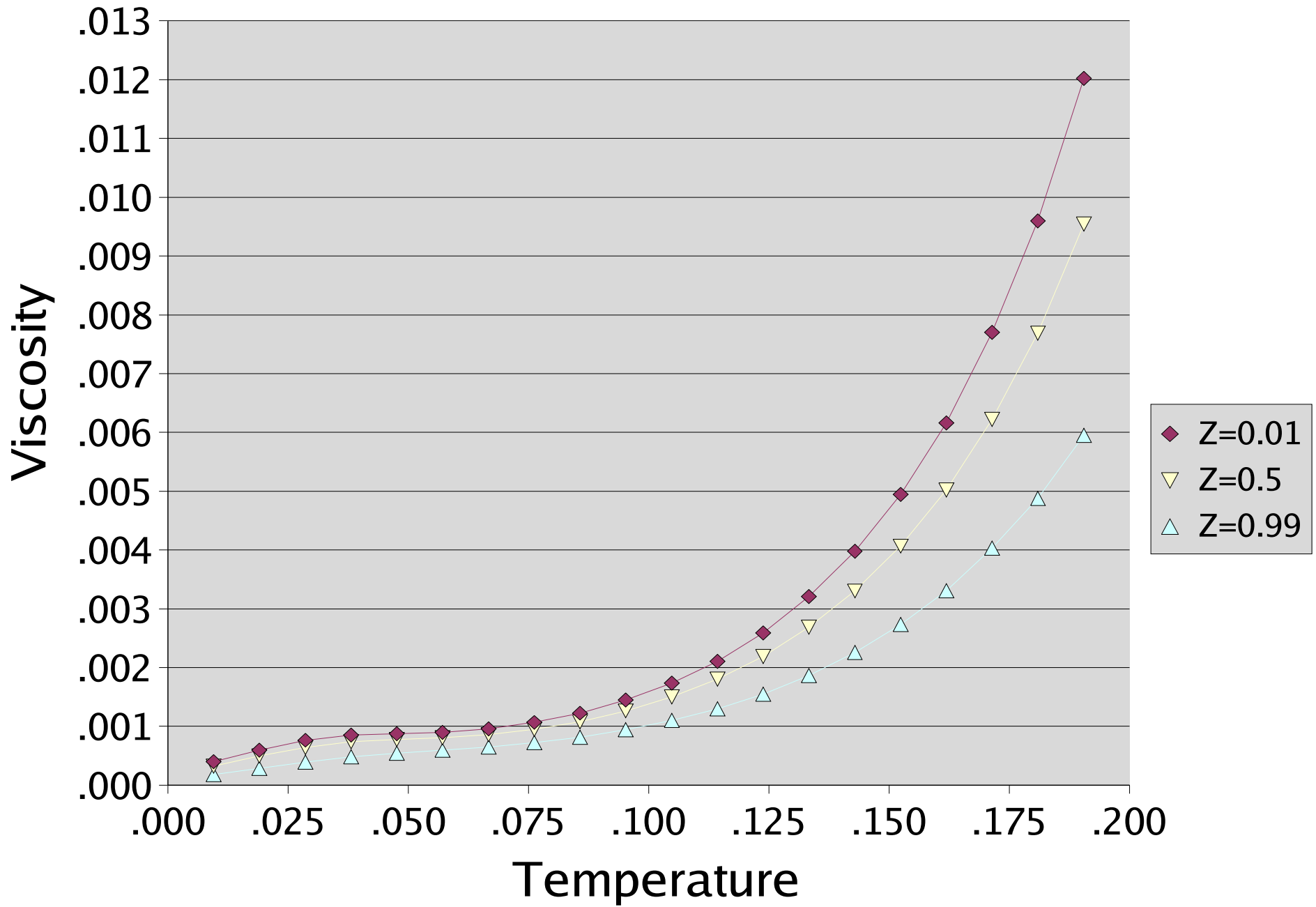

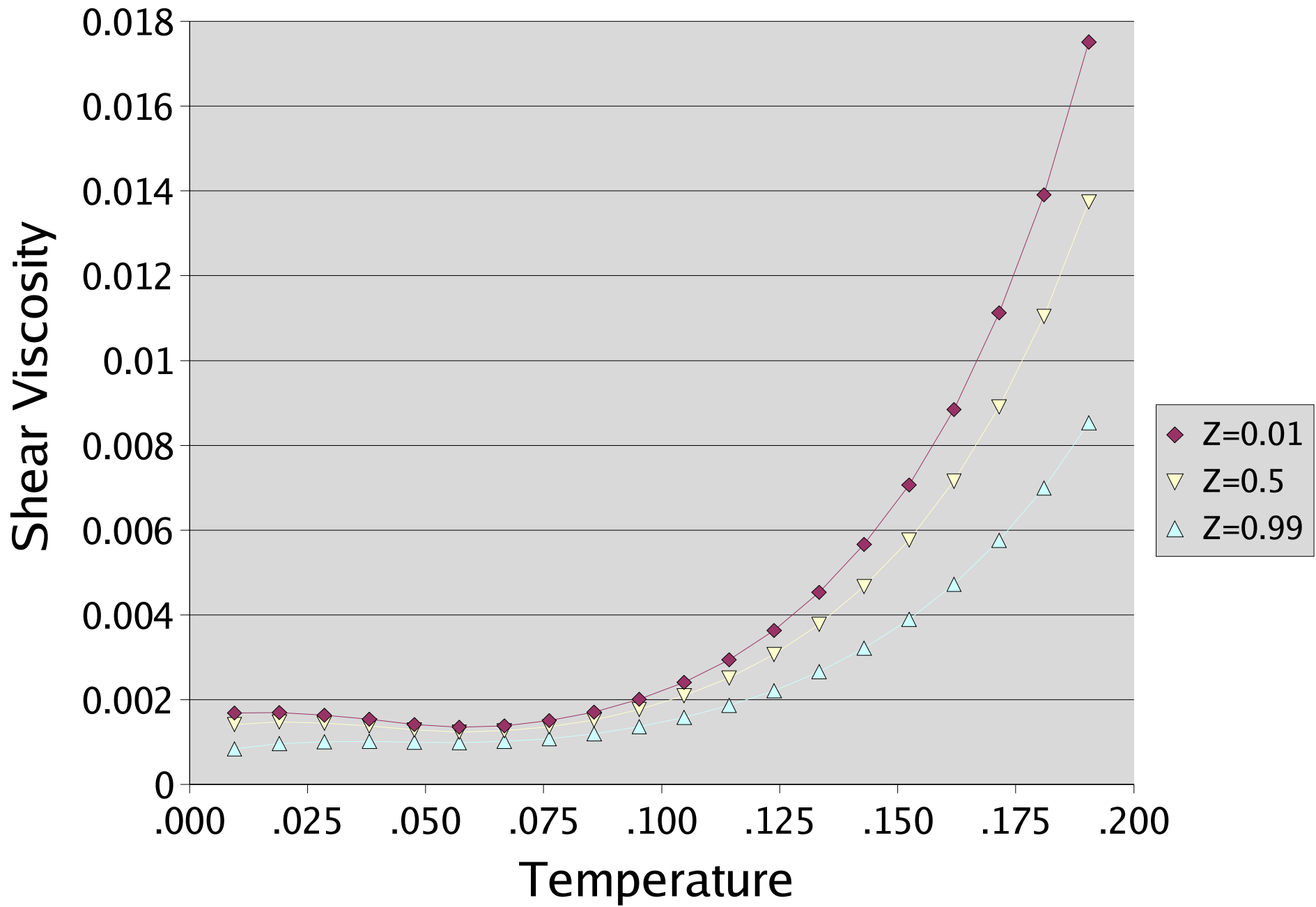